\documentclass[twocolumn]{aastex63}
\usepackage[utf8]{inputenc}
\usepackage{subfigure}
\usepackage{amsmath,amssymb,amsfonts}
\usepackage{color}

\newcommand{\pp}[1]{\textcolor{blue}{ #1}}

\begin{document}

\title{Synchronizing the EMRIs and IMRIs in AGN accretion disks}

\author{Peng Peng}
\affiliation{Astronomy Department, School of Physics, Peking University, 100871 Beijing, China}

\author{Xian Chen}
\correspondingauthor{Xian Chen}
\email{xian.chen@pku.edu.cn}
\affiliation{Astronomy Department, School of Physics, Peking University, 100871 Beijing, China}
\affiliation{Kavli Institute for Astronomy and Astrophysics at Peking University, 100871 Beijing, China}

\begin{abstract}
Extreme-mass-ratio inspirals (EMRIs) and intermediate-mass-ratio inspirals
(IMRIs) are important gravitational-wave (GW) sources for the Laser
Interferometer Space Antenna (LISA).  So far, their formation and evolution are
considered to be independent, but recent theories suggest that stellar-mass
black holes (sBHs) and intermediate-mass black hole (IMBHs) can coexist in the
accretion disk of an active galactic nucleus (AGN), which indicates that EMRIs
and IMRIs may form in the same place.  Motivated by the fact that a gas giant
migrating in a protoplanetary disk could trap planetesimals close to its orbit,
we study in this paper a similar interaction between a gap-opening IMBH in an
AGN disk and the sBHs surrounding it. We analyse the torques imposed on the
sBHs by the disk as well as by the IMBH, and show that the sBHs can be trapped by
the IMBH if they are inside the orbit of the IMBH.  Then we implement the
torques in our numerical simulations to study the migration of an outer IMBH
and an inner sBH, both embedded in an AGN disk.  We find that their migration
is synchronized until they reach a distance of about ten Schwarzschild radii
from the central supermassive black hole, where the pair breaks up due to
strong GW radiation.  This result indicates that LISA may detect
an EMRI and an IMRI within several years from the same AGN. Such a GW source will bring
rich information about the formation and evolution of sBHs and IMBHs in AGNs.
\end{abstract}


\section{Introduction}\label{sec:intro}

Extreme-mass-ratio inspirals (EMRIs) and intermediate-mass-ratio inspirals
(IMRIs) are important gravitational-wave (GW) sources for milli-Hertz (mHz) GW
detectors, such as the Laser Interferometer Space Antenna
\citep[LISA,][]{pau18}.  An EMRI normally consists of a supermassive black hole
(SMBH) of $10^6-10^7$ $M_\odot$ and a stellar-mass black hole (sBH) of tens of
solar masses, so that the mass ratio $q$ is much smaller than $10^{-4}$.  An
IMRI has a typical mass ratio of $10^{-4}\lesssim q\lesssim10^{-3}$
\citep{2007CQGra..24R.113A}. Therefore, it can be formed either by an
intermediate-mass black hole (IMBH) of $10^2-10^4$ $M_\odot$ and a stellar-mass
compact object (i.e., a white dwarf, neutron star, or sBH), or by an SMBH and an
IMBH.  Since the GW radiation timescale is proportional to $q^{-1}$, EMRIs and
IMRIs could dwell in the LISA band for many years. During this period, they
accumulate $10^4-10^5$ GW cycles in the data, providing rich information about
the spacetime geometry close to the horizon of a massive black hole
\citep{2007CQGra..24R.113A}.

One important place for EMRI and IMRI formation is the accretion disk of an
active galactic nucleus (AGN).  It is known to be the breeding ground of sBHs
\citep[e.g.][]{syer91, artymowicz93, subr99, karas01, levin03, Goodman2004}.
The sBHs could form from the massive stars in the outer part of the disk
\citep{Goodman2004}, or be captured from the nuclear star cluster by the disk
\citep{syer91}.  Once inside the disk, the sBHs could migrate towards the
central SMBH due to various hydrodynamical effects, including hydrodynamical
drag \citep{sigl07}, the interaction with density waves \citep{levin03,
kocsis11, kocsis12, sanchez20}, and head or tail winds if the disk is
geometrically thick \citep{chakrabarti93, molteni94, basu08, kocsis11}.  Those
sBHs eventually reaching a few Schwarzschild radii from the central SMBH will
produce EMRIs.  Such ``wet'' EMRIs may predominate the event rate as recent
calculations suggest \citep{grobner20, pan21, pan21b, 2022arXiv220510382D}. 

The accretion disk of an AGN may also contain IMBHs, and hence could produce
IMRIs as well. On one hand, IMBHs can be brought into the galactic nucleus by
galaxy mergers or inspiraling globular clusters \pp{\citep{2014ApJ...796...40M,
volonteri03}}.  These IMBHs later could be captured by the accretion disk in a
way similar to the capture of sBHs \citep{Ivanov1999}.  On the other hand,
IMBHs could also grow from the aforementioned sBHs accumulated in the AGN disk.
The growth is due to either the accretion of the surrounding gas
\citep{mckernan11,2012MNRAS.425..460M,2021MNRAS.507.3362T} or the merger with
the other sBHs in the disk \citep{mckernan12, bartos17, stone17}.

Earlier studies have shown that the probability of IMBH formation could be
particularly high at a series of radii in the AGN disk, known as the
``migration traps'' \citep{bellovary16,secunda19,secunda20, peng21}.  Such
traps typically reside at tens to hundreds of Schwarzschild radii ($R_{\rm S}$)
from the central SMBH.  Inside the migration trap, the hydrodynamical torque
which drives the migration of sBHs vanishes. Consequently, sBHs will accumulate
in the trap, forming binaries. The binaries will continue to harden due to the
interaction with either the surrounding gas \citep{baruteau11,antoni19} or the
other sBHs \citep{leigh18, yang19}.  A likely outcome is that the binaries will
merge within the lifetime of the AGN. Moreover, the merger remnants will remain
in the accretion disk because the recoil velocity (due to anisotropic GW
radiation) is typically much smaller than the escape velocity from the central
SMBH \citep{miller09a,2019PhRvL.123r1101Y,2021MNRAS.507.3362T}.  Therefore, the
post-merger sBHs could participate in the next generation of mergers.  Through
such ``hierarchical mergers'' \citep{2019PhRvL.123r1101Y,2021NatAs...5..749G},
sBHs could gradually grow into the intermediate-mass range.  Interestingly, the
above channel of heavy black hole (BH) formation in AGN disk can well explain
the merger rate  \citep{mckernan18, grobner20, secunda20} and the mass function
\citep{gayathri20} of the binary BHs detected by the Laser Interferometer
Gravitational-wave Observatory (LIGO) and the Virgo detectors.

As the IMBH in an AGN disk grows, the tidal torque induced by the IMBH on the
surrounding gas would eventually be strong enough to open a gap in the disk
\citep{ivanov99, gould00}. This gap-opening process is similar to what is
happening around the gas giants in protoplanetary disks
\citep{1979MNRAS.186..799L, 1979MNRAS.188..191L,lin86, artymowicz94}. After
opening a gap, the IMBH could continue exchanging energy and angular momentum
with the accretion disk by its tidal force, and hence keep migrating towards
the central SMBH \citep{gould00, armitage02}. 

According to the above picture of the formation and evolution of sBHs and IMBHs
in an AGN accretion disk, it is likely that an IMBH could encounter other sBHs
during its migration.  For example, the IMBH could catch up with an sBH if the
migration of the latter is slower, or the IMBH may at some point pass by a
migration trap of sBHs.  The subsequent interaction seems analogous to the
interaction between a gap-opening gas giant and a sub-gas-opening terrestrial
planet in a protoplanetary disk.  Previous studies of such an interaction
suggest that the smaller terrestrial planet could be trapped in one of the
resonance belts around the larger gas giant
\citep{1996LPI....27..479H,2008MNRAS.386.1347P, 2008A&A...482..333P}.  

For the same reason, we also expect a gap-opening IMBH in an AGN disk to trap
an sBH along its way towards the central SMBH. If such a pair could be
maintained until the IMBH reaches the last few Schwarzschild radii from the
central SMBH, we would detect an interesting event--an EMRI happening at the
same time of an IMRI.  Such a GW source could reveal rich triple dynamics in
the regime of strong gravity \citep[see, e.g.,][for the interaction between two
EMRIs]{2021PhRvD.104d4056G}. In fact, \cite{Yang2019} have shown that two
sub-gap-opening sBHs in the accretion disk of an AGN could indeed be trapped at
the mean-motion resonance and migrate, as a pair, towards smaller radius.
However, they also showed that the pair will decouple when the inner sBH
reaches tens of $R_{\rm S}$, where GW radiation of the inner EMRI becomes
important.  

The situation could be different if an IMBH traps an sBH within its orbit.
First, the coupling between the sBH and the IMBH is expected to be tighter than
an sBH-sBH pair, because the IMBH exerts a larger tidal torque and excites
stronger density waves in the disk.  Second, the inspiral timescale of the IMBH
due to GW radiation could be shorter than that of the sBH, so that the IMBH
could keep up with the inward migration of the sBH even when the latter has
entered the GW-radiation regime.  To test these postulations, we conduct in
this paper numerical simulations of the migration and interaction of the IMBHs
and sBHs in an AGN accretion disk.

The paper is organized as follows. In Section~\ref{sec:single}, we review the
physical processes which  will lead to the formation and migration of sBHs and
IMBHs in the accretion disk of an AGN. We show that a gap-opening IMBH can
catch up with the migration of an inner sBH. In Section~\ref{sec:trapping}, we
calculate the hydrodynamical torque of the gas and the tidal torque of the IMBH
exerted on the sBH. Based on the understanding of the torques, we conduct 1D
and 2D simulations in Section~\ref{sec:synchronization} to show how the IMBH
and the sBH are synchronized in their migration towards the central SMBH.
Finally, in Section~\ref{sec:conclusion} we discuss the possibility of
detecting an EMRI-IMRI pair by the LISA mission and the relevant parameter
space which can produce such a special GW source.

\section{Formation and migration of sBHs and IMBHs in AGN disk \label{sec:single}}

Previous studies suggest that sBHs can be produced in the accretion disks of
AGNs in two ways. First, sBHs can originate from the nuclear star cluster
surrounding an AGN. If their orbits intersect the accretion disk, the repeated
collision with the disk could cause energy and angular-momentum loss, and some
sBHs could eventually be captured by the disk \citep{syer91}. In particular,
\citet{2020MNRAS.499.2608F} and \citet{2022arXiv220709540N} showed that about
$10 \%$ of the sBHs with a semimajor axis of $a_s \sim$ $10^3$ -- $10^6$
$R_{\rm{S}}$ relative to the SMBH can be captured from the nuclear star cluster
into the accretion disk during the lifetime of the AGN.  In the case of an SMBH
with a mass of $M_{\rm{SMBH}}=10^6 M_\odot$, about $100$ sBHs can be captured
because there are about $1000$ sBHs in the nuclear star cluster whose orbits
intersect the accretion disk \citep[see, e.g., Section 5.5 in][]{tagawa20}.
Second, sBHs could be produced by the massive stars in the accretion disk.
These stars are either born in the disk \citep{Goodman2004, 1987Natur.329..810S} or captured from the nuclear star cluster
\citep{artymowicz93}. However, recent calculations suggest that this channel is
not dominating the formation of sBHs in AGN disks, since only a small fraction
of the stars ($\sim 1\%$) would evolve into sBHs \citep{tagawa20}.  Therefore,
in the following analysis we focus on the sBHs captured from the nuclear star
cluster.

A captured sBH will excite density
waves in the accretion disk \citep{1978ApJ...222..850G, 1979ApJ...233..857G}. 
The back reaction of the density waves effectively exerts a torque on the sBH,
causing the sBH to migrate in the disk, which is known as the ``Type-I migration''
\citep{1980ApJ...241..425G}. 
The corresponding
migration timescale can be calculated with
\begin{align}
    & \nonumber T_I = \frac{f_1 h^2 M_{\rm{SMBH}}^2}{m_{\rm{sBH}} \Sigma a_s^2 \Omega_s} \\
    & \nonumber = 10^4 f_1 \left(\frac{m_{\rm{sBH}}}{10 M_\odot}\right)^{-1} 
	\left(\frac{M_{\rm{SMBH}}}{10^{6}M_\odot}\right) \left(\frac{h}{10^{-2}}\right)^{2} \\
    & \times \left(\frac{\Sigma}{10^6 \, {\rm{g}} \, {\rm{cm}}^{-2}}\right)
	\left(\frac{a_s}{1000 R_{\rm{S}}}\right)^{-1/2}\,{\rm yr}, 
	\label{eq:T_I}
\end{align}
where $m_{\rm{sBH}}$ is the mass of the sBH,  $\Omega_s$ is its angular
velocity, $h$ and $\Sigma$ are, respectively, the aspect ratio (between the
scale height and the radius) and the surface density of the accretion disk, and
$f_1$ is a function of the temperature and density gradients of the disk near
the sBH \citep[e.g.][]{ Paardekooper11}. Given the parameters of our interest,
we find that the Type-I migration timescale is much shorter than the lifetime
of the AGN \citep[$\sim 10$ Myr,][]{2008ApJ...676..816G, 2009ApJ...698.1550H,
2013MNRAS.434..606G}.

The direction of the Type-I torque depends on the relative strength of the
density waves leading and trailing the sBH. It is characterized by the
parameter $f_1$ in Equation~(\ref{eq:T_I}).  In a large range of radius, the
trailing wave is stronger so that the torque on the sBH is negative. Therefore,
the sBH loses angular momentum and migrates inward, towards the central SMBH.
However, there are regions in the disk where the temperature and density
profiles are discontinuous due to the change of opacity. In such a region, the
leading wave could become stronger and the sBH will feel a positive torque and
migrate outward \citep{bellovary16}. The transition from a negative torque to a
positive one will result in a zero-torque region.  Here the inwardly migrating
sBHs will meet the outwardly migrating ones, and the migration will stall once
the sBHs enter the region. For this reason, the region with zero torque is
known as the ``migration trap'', and there are a series of such traps at a
radial range between several tens and several thousands of $R_{\rm{S}}$
\citep{bellovary16, secunda19}.

Inside a migration trap, the sBHs, assisted by the surrounding gas, could
experience multiple mergers. The number of mergers depends on the number of
sBHs in the trap.  \cite{secunda20} recently estimated that for a SMBH of
$M_{\rm SMBH} = 10^8 M_\odot$, every $\sim10^5$ years an sBH will be
transported to the Type-I migration trap.  Consequently, a total number of
$\sim$ 100 sBHs will end up in the trap during the AGN lifetime.  This result
suggests that an IMBH with a final mass of $\sim$ $1000 M_{\odot}$ could form
\citep[see Figure~4 in][]{secunda20}.  For a smaller SMBH, although there
are less sBHs in the nuclear star cluster \citep[see, e.g., Section 5.5
in][]{tagawa20}, the capture of sBHs by the disk could be more efficient
because the dynamical timescale is shorter. In addition, the Type-I migration
timescale also decreases with decreasing $M_{\rm SMBH}$.  Therefore, the number
of sBHs transported to the migration trap, as well as the final mass of the
IMBH, could be even larger. 

Besides forming in the migration trap, the IMBHs in AGN disks could come from
other channels.  For example, \citet{mckernan12} and \citet{tagawa20} showed
that an IMBH of $100 M_{\odot}$ could form outside the migration trap, partly
due to mutual capture of sBHs and partly to gas accretion. Moreover, galaxy
mergers and inspiraling globular clusters could also bring in IMBHs, as has
been mentioned in Section~\ref{sec:intro}. 

Therefore, IMBHs and sBHs are likely to co-exist in an AGN accretion disk.  To
understand the subsequent evolution, it is important to realize that their
migration could be different. While a sBH undergoes Type-I migration, an IMBH
could open a gap in the accretion disk which characteristically changes the
subsequent evolution.  The criterion of opening a gap is determined by two
factors.  First, the tidal torque of the IMBH should be stronger than the
viscous torque of the disk, which gives 
\begin{equation}
    \frac{(m_{\rm{IMBH}}/M_{\rm{SMBH}})^{1/3}}{h} (\frac{ h}{1600  \alpha})^{1/3} > 1 \, 
    \label{eq:gap_1}
\end{equation}
\citep{1986ApJ...309..846L},
where $m_{\rm{IMBH}}$ is the mass of the IMBH and $\alpha$ is the viscosity
parameter of the disk. Second, the tidal force of the IMBH should be
stronger than the pressure at the edge of the gap, which gives an additional condition
\begin{equation}
    \frac{(m_{\rm{IMBH}}/M_{\rm{SMBH}})^{1/3}}{h} > 1 \, 
    \label{eq:gap_2}
\end{equation}
\citep{1997Icar..126..261W}.  So for typical parameters, $ M_{\rm{SMBH}} = 10^6
M_\odot$,  $\alpha = 0.01$, and $h \lesssim 0.05$, the IMBH needs to be more
massive than $\sim 100 M_\odot$ to open a gap.  This mass corresponds to $\sim
10$ mergers of sBHs. This number is smaller than the typical number of sBHs in
an migration trap.   

A gap-opening IMBH is no longer subject to the Type-I torque because the
density profile of the disk is drastically altered.  Instead, the IMBH is
coupled to the viscous evolution of the disk.  It exchanges energy and angular
momentum with the disk by tidally interacting with the gas at the edges of the
gap.  The corresponding migration is known as the ``Type-II migration''
\citep{lin86}. The timescale can be calculated with
\begin{align}
    & \nonumber T_{II} = \frac{1}{\alpha h^2 \Omega_I} = 10^4  
	\left(\frac{M_{\rm{SMBH}}}{10^{6}M_\odot}\right) \left( \frac{a_I}{1000 R_{\rm{S}} } \right)^{3/2}  \\ 
	& \times \left(\frac{h}{10^{-2}}\right)^{-2}\left(\frac{\alpha}{10^{-2}}\right)^{-1} {\rm yr},
	\label{eq:T_II}
\end{align}
where $a_I$ and $\Omega_I$ are the semimajor axis and the orbital angular
velocity of the IMBH.  The direction of the migration may depend on the mass
ratio between the IMBH and the central SMBH \citep{2017MNRAS.466.1170M,
2020ApJ...889..114M}.  But for the IMBHs of our interest ($m_{\rm
IMBH}\lesssim10^3M_\odot$), the mass ratio is small and the migration is
normally towards the SMBH.

Equations~(\ref{eq:T_I}) indicates that the migration of sBHs will slow down as
they approach the SMBH, since $T_{I}\propto a_s^{-1/2}$. The migration may even
stall if the sBHs enter one of the Type-I migration traps. For IMBHs, however,
Equations~(\ref{eq:T_II}) suggests that their Type-II migration will accelerate
since $T_{I}\propto a_I^{3/2}$. Such a differential migration will result in an
interesting and important consequence: an outer IMBH can catch up with an inner
sBH.   

\section{Interaction between sBHs and IMBHs \label{sec:trapping}}

When an outer IMBH catches an inner sBH, two factors will affect the subsequent
evolution.  First, the sBH will start interacting with the inner edge of the
gap opened by the IMBH. Since the edge of the gap has a sharp density and
temperature profile, it will  modify the Type-I torque on the sBH.  Second, the
IMBH could exchange energy and angular momentum with the sBH if the two BHs are
caught in a resonance. We will analysis these two effects in this section and
show that the migration of the two BHs can be synchronized if the sBH is inside
the orbit of the IMBH.

\subsection{Modified Type-I torque due to the gap}

\begin{figure}
\centering
\includegraphics[width=0.48\textwidth]{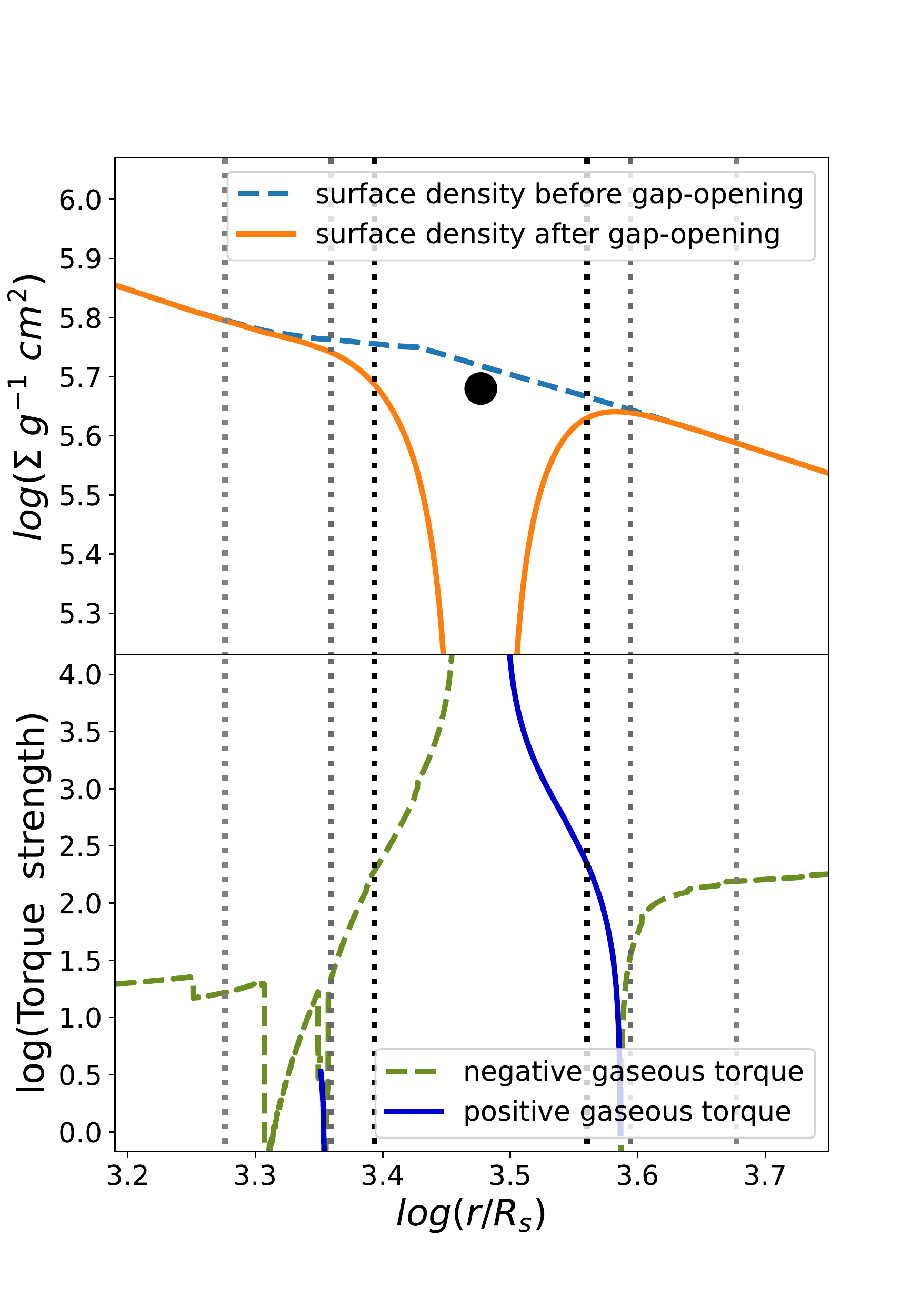}
\caption{ \textbf{Upper panel}: Surface density of the disk before (blue dashed curve)
and after (orange solid curve) the gap is opened. 
The black dot marks the location of the IMBH. 
The vertical dotted
lines, from left to right, show the loci of the 2:1, 3:2, 4:3,
3:4, 2:3, and 1:2 resonances. 
\textbf{Lower panel}: Strength of the Type-I torque (in arbitrary unit) near the gap. 
The green dashed line shows the negative torque, which drives a
sBH migrating inward, and the blue solid one shows the positive torque,
driving the sBH outwards.} \label{fig:sketch} 
\end{figure}

To evaluate the first effect, we use the 1D hydrodynamical code presented in
\citet{Fontecilla19} to calculate the surface-density profile of the gas near
the gap opened by the IMBH.  The result is shown in the upper panel of
Figure~\ref{fig:sketch}.  The parameters are $m_{\rm{IMBH}} = 100 M_\odot$,
$M_{\rm{SMBH}} = 10^6 M_\odot$, and the accretion rate of the disk
$\dot{M}_{\rm{SMBH}}$ is set to the Eddington rate $\dot{M}_{\rm{Edd}} = L_{\rm{Edd}} / ( 0.1 c^2 )$. We can see that near the gap the
surface density decreases by several orders of magnitude within a radial range
of only $0.1$ dex. 

The sharp change of the density, as well as the temperature, will significantly
affect the value and the sign of $f_1$ in Equation~(\ref{eq:T_I}).  Using the
formulae given in \cite{2010MNRAS.401.1950P} to calculate $f_1$, we can
calculate the strength of the Type-I torque at the edges of the gap. The result
is shown in the lower panel of Figure~\ref{fig:sketch}.  We find that the
Type-I torque at the inner (outer) edge of the gap is negative (positive). This
result indicates that the torque tends to ``push'' sBHs away from the gap.
Moreover, the magnitude of the torque increases sharply towards the gap,
indicating that it is difficult for sBHs to enter the gap.

\subsection{Mean-motion resonance}

The second effect is the gravitational interaction between the IMBH and the
sBH.  The interaction can be modeled by the perturbation theory
\citep{2000ApJ...535..385F} developed for planetary systems.  Such a theory is
applicable in our case because both the IMBH and the sBH are much lighter than
the central SMBH, and their orbital eccentricities and inclinations remain
small due to a hydrodynamical damping effect \citep{2004ApJ...602..388T}.
According to this theory, the interaction is the strongest at the mean-motion
resonances, where the orbital frequencies of the IMBH and the sBH take integer
ratios, such as  $\Omega_s/\Omega_I = 1:2,\, 2:1,\, 3:4,\, etc $
\citep{1999ssd..book.....M}.  The loci of the most important resonances are
shown in the Figure~\ref{fig:sketch} as the dotted vertical
lines. 

The resonant interaction exchanges energy and angular momentum between the IMBH
and the sBH, so the semimajor axis $a_s$ and the orbital eccentricity $e_s$ of
the sBH will secularly change. We will ignore the back reaction on the
evolution of the IMBH since the mass of the sBH is small.  Take the 2:1
resonance for example, i.e., $\Omega_s/\Omega_I = 2$ and the sBH is inside the
orbit of the IMBH. The evolution is governed by the equations
\begin{align}
	& \frac{\dot{a}_s}{a_s} = {\rm sgn}(\Omega_s-\Omega_I) \, f_d \,  e_s \, \Omega_s \, \sin(\varphi)
\left(\frac{4\,m_{\rm{IMBH}}}{M_{\rm{SMBH}}} \right)
	\label{eq:resonance1}, \\
    & \dot{e}_s = -f_d \, \Omega_s \, \sin(\varphi)\left(\frac{m_{\rm{IMBH}}}{M_{\rm{SMBH}}} \right)
, \label{eq:resonance2} \\
    & \dot{\varphi} = - \Omega_s + 2 \Omega_I 
    -f_d \, e_s^{-1} \, \Omega_s \, \cos(\varphi)\left(\frac{m_{\rm{IMBH}}}{M_{\rm{SMBH}}} \right)
    \label{eq:resonance3}
\end{align}
\citep{1999ssd..book.....M}, where a dot denotes the time derivative, $f_d$ is
a factor of order unity which takes different values for different resonances,
and $\varphi$ is the resonant argument. In Equation~(\ref{eq:resonance3}) we
have omitted the higher-order terms of $e_s$ and $m_{\rm{IMBH}}/M_{\rm{SMBH}}$.
We note that $f_d$ is different from $f_1$, and it is negative when
$\Omega_s>\Omega_I$.  Besides the 2:1 resonance, we have also computed the 3:4,
4:5, and 5:6 resonances to the first order of $e_s$ and included them in our
model.  The other resonances of higher order of $e_s$ are neglected in this
work since $e_s$ remains small (see below).  In the case where the sBH is
outside the orbit of the IMBH, we have $\Omega_s < \Omega_I$, and the factor
$f_d$ in Equation~(\ref{eq:resonance1}) will be positive.

The last three equations indicate that the sBH could be trapped by the IMBH in
one of the mean-motion resonances when $\Omega_s>\Omega_I$.  For example,
Equation~(\ref{eq:resonance3}) suggests that $\varphi$ evolves on a timescale
much longer than the orbital period when the sBH resides at the exact location
of the resonance $\Omega_s /\Omega_I=2$. This location is where $\varphi=0$.
If the sBH deviates from this location and moves, say, further inward
(outward), then $- \Omega_s + 2 \Omega_I$ will become negative (positive),
making $\varphi$ negative (positive) as well. Consequently, in
Equation~(\ref{eq:resonance1}) $\dot{a}_s$ will be positive (negative) since
$f_d<0$ when $\Omega_s>\Omega_I$.  Therefore, the sBH feels a positive
(negative) torque and will move back into the resonance.  For this reason, the
sBH will be trapped at a location where $\varphi\simeq0$ and $\dot{a}\simeq0$.
In fact, when $\Omega_s<\Omega_I$, i.e., the sBH is outside the orbit of the
IMBH, the equations for resonant interaction (not shown here) will lead to the
same result.

The above physical picture of a resonance trap needs to be modified in our
problem because (i) the IMBH migrates and (ii) gas is present around the sBH.
The consequences are twofold.  On one hand, to keep up with the inward Type-II
migration of the IMBH, the sBH should satisfy the condition $\dot{a}_s<0$. Such
a requirement, according to Equation~(\ref{eq:resonance1}), corresponds to the
condition $\varphi>0$ for either $\Omega_s>\Omega_I$ or $\Omega_s<\Omega_I$.
Under this condition and  according to Equation~(\ref{eq:resonance2}), the
eccentricity of the sBH, $e_s$, will increase (decrease) when
$\Omega_s>\Omega_I$ ($\Omega_s<\Omega_I$).  On the other hand, gas tends to
damp the eccentricity $e_s$ \citep{2004ApJ...602..388T}. This effect could
counterbalance the increase of $e_s$ when $\Omega_s>\Omega_I$, but will further
reduce $e_s$ when $\Omega_s<\Omega_I$.  In the former case, i.e., the sBH is
inside the orbit of the IMBH, the balance will result in a synchronized
migration of the sBH and the IMBH.  In the later one, however, $e_s$ will
quickly diminish. Then the resonance torques disappear because they are
proportional to $e_s$, and the sBH can no longer keep up with the migration of
the IMBH.

\section{Synchronizing EMRIs and IMRIs \label{sec:synchronization}}

We have seen that the evolution of the sBHs and IMBHs in an AGN disk is
determined by a complex interplay between the Type-I/Type-II migration, the
excitation/reduction of $e_s$ by mean-motion resonances, and the hydrodynamical
damping of $e_s$.  To understand the long-term evolution, we conduct 1D and 2D
hydrodynamical simulations in this section.

\subsection{Early evolution at large radius}
\label{subsec:early}

We first use the 1D hydrodynamic code developed in \cite{Fontecilla19} to simulate
the gap opening process and the subsequent Type-II migration of an IMBH in an
AGN disk.  The initial condition of the disk is derived from the model
described in \cite{Goodman2003}, which results in multiple Type-I migration
traps in the disk.  In our fiducial model, we choose the parameters $
M_{\rm{SMBH}} = 10^6 M_\odot$, $\dot{M}_{\rm{SMBH}} = \dot{M}_{\rm{Edd}}$,
$\alpha=0.01$, and $m_{\rm{IMBH}} = 100 M_\odot$.  

The initial location of the IMBH is $a_I=3000 R_{\rm{S}}$, which coincides with
a Type-I migration trap in the unperturbed disk. This choice is motivated by
the prediction that an IMBH could be produced inside a Type-I migration trap (see
Section~\ref{sec:single}).  However, if the IMBH comes from a capture event as
is mentioned in Section~\ref{sec:intro}, it could also migrate to the first
Type-I migration trap and interact with the sBHs there. In this latter case,
our choice of $a_I$ is also reasonable.
 
For sBHs, initially we place one at $a_s=0.8 a_I$ and another at $ = 1.2 a_I$,
to mimic the result found by earlier hydrodynamical simulations of the Type-I
migration trap \citep{secunda19, secunda20}.  To compute the speed of migration,
we consider both the Type-I torque and the torque due to mean-motion resonance.
(i) The Type-I torque is calculated according to the equations in
\cite{Paardekooper11}.  In the calculation, the disk profile, in particular the
profiles at the edges of the gap, is given by our 1D simulation.  (ii) For the
resonance torque, we notice that in our simulations the sBHs do not enter the
annulus between the 6:7 and 7:6 resonances. Therefore, we only consider the
2:1, 3:4, 4:5, 5:6, 1:2, 4:3, 5:4, and 6:5 resonances.  The corresponding
equations governing the evolution of $a_s$ and $e_s$ are adopted from
\cite{1999ssd..book.....M}. 

For the two sBHs, we also consider the hydrodynamical damping of their orbital
eccentricities. We model the effect by a characteristic damping timescale,
$T_d\equiv-e_s/\dot{e}_s$. When the eccentricity is smaller than
the aspect ratio of the disk, i.e., when $e_s \lesssim h$, we use  $T_d=h^2 T_{I}$ to calculate the damping timescale  \citep{2004ApJ...602..388T}.
Correspondingly, the damping rate, $-e_s/T_d$, is
proportional to $e_s$.  When $e_s > h$, however, the damping rate will decrease
as $e_s$ increases \citep[e.g.][]{2000MNRAS.315..823P}. In this case, the
evolution of the sBHs is more complex, and we will study it in more detail in
the next subsection.

\begin{figure}
\centering
\includegraphics[width=0.5\textwidth]{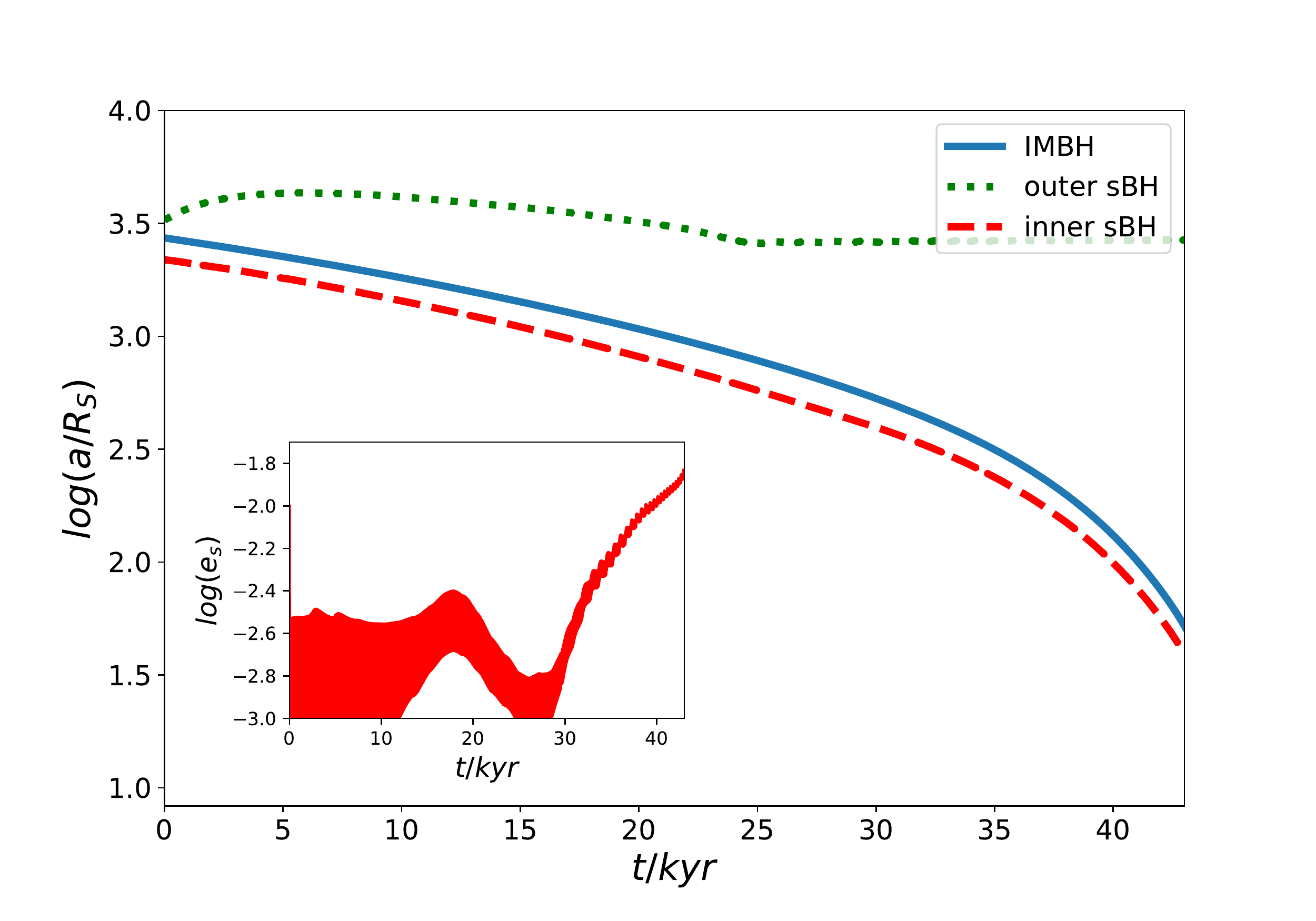}
\caption{Migration of an IMBH and two sBHs 
embedded in the accretion disk of an AGN.
The blue solid line shows the semimajor axis of the IMBH as a function of time,
and the red dashed and the green dotted lines refer to the two sBHs,
one inside the orbit of the IMBH and the other outside.
The inset shows the orbital eccentricity of the
inner sBH as a function of time.} \label{fig:migration} 
\end{figure}

Figure~\ref{fig:migration} shows the early evolution of the IMBH and the two
sBHs when the condition $e_s\lesssim h$ can be satisfied.  We can see that the
sBH inside the orbit of the IMBH remains trapped during the Type-II migration
of the IMBH. The migration of the two BHs are ``synchronized''.  The other sBH
outside the IMBH, however, is initially pushed away from the IMBH.  This
behavior can be explained by the positive Type-I torque at the outer edge of
the gap.  Moreover, this outer sBH cannot keep up with the Type-II migration of
the IMBH, a result consistent with the prediction made in the previous section
based on a qualitative analysis of the resonances.  As the IMBH leaves its
original location, we observe that the outer sBH stays more or less at the same
initial radius. This is caused by a restore of the disk to the unperturbed
state and a recovery of the Type-I migration trap at the location of the outer
sBH.

We find that the coupling between the IMBH and the inner sBH is facilitated by
different mechanisms at different evolutionary stages.  (i) At the beginning,
the IMBH migrates relatively slowly.  The negative Type-I torque at the inner
edge of the gap (see Figure~\ref{fig:sketch}) is sufficient to drive the sBH
inward, at the same pace with the IMBH.  The torque due to mean-motion
resonance is subdominant at this stage, since the orbital eccentricity of the
sBH remains small (see the inset of Figure~\ref{fig:migration}).  (ii) When the
IMBH reaches a semimajor axis of $a_I\sim 10^2R_{\rm{S}}$, the migration
becomes relatively fast since $T_{II}\propto a_I^{3/2}$. At the same time, the
aspect ratio of the disk, $h$, has increased significantly, making the Type-I
torque less effective ($T_I\propto h^2$).  As a result, the torque due to
mean-motion resonance starts to dominate. Since the resonance torque is
proportional to $e_s$, as Equation~(\ref{eq:resonance1}) suggests, the sBH
needs to maintain a relatively high eccentricity to keep pace with the IMBH.
For this reason, in the inset of Figure~\ref{fig:migration}, we see a fast rise
of $e_s$ towards the end of the simulation.  In fact, the sBH is trapped in the
3:2 resonance at this stage because we find that the ratio $a_s/a_I$ converges
to $(2/3)^{2/3}\simeq 0.76$. 

\subsection{Late evolution in the GW regime}
\label{subsec:later}

When the IMBH migrates to a distance of $a_I\sim30R_{\rm{S}}$ from the central
SMBH, GW radiation becomes important and hence the IMBH migrates even faster.
A direct consequence is that $e_s$ is excited to a value higher than the aspect
ratio $h$ of the disk.  At this stage, we can no longer calculate the
hydrodynamical damping timescale of the eccentricity  by $h^2 T_{I}$.

To properly calculate the damping timescale for $e_s > h$, we use the
relationship $T_d=f_e e_s^3$ found in the previous studies on the migration of
planetesimals with high eccentricities \citep[e.g.][]{ 2000MNRAS.315..823P,
2007A&A...473..329C, 2008A&A...482..677C, 2010A&A...523A..30B,
2011ApJ...737...37M, 2020MNRAS.494.5666I}.  The coefficient $f_e$ is determined
with the following considerations. By comparing Equation~(32) in
\citet{2000MNRAS.315..823P} with the empirical fitting formula in
\citep{2008A&A...482..677C}, as well as the formula derived from a
dynamical-friction model \citep{2011ApJ...737...37M, 2020MNRAS.494.5666I}, we
find a discrepancy of factor of two for $f_e$.  Therefore, we use the value
derived in \citet{2000MNRAS.315..823P} with the default softening parameter as
our default $f_e$, and increase it by a factor of three to account for the
current theoretical uncertainties. The resulting $T_d$ is longer than that for
$e_s < h$. 

Besides a longer damping timescale, the interaction between the IMBH and the
inner sBH is also characteristically different. Now the orbital eccentricity of
the sBH is high as we have seen in Figure~\ref{fig:migration}.  The resonance
model based on perturbation theory and presented in the previous subsection
becomes invalid.  Therefore, we drop the 1D code and use, instead, the 2D
N-body simulation package {\tt rebound} \citep{rebound, reboundias15} to study
the gravitational interaction between the two BHs when $a_I<30R_{\rm{S}}$.
Since {\tt rebound} does not include hydrodynamical effects, we implement an
extra force term to the sBH to account for the hydrodynamical damping of $e_s$
on the timescale of $T_d$ \citep[following][]{2014MNRAS.445..479C}.  In
addition, to include the effect of GW radiation, we imposed an extra friction
on the IMBH as well as the sBH.  The magnitude of this frictional force is
calculated from the energy-loss timescale due to GW radiation \citep{peters63}.

\begin{figure}
    \centering
    \includegraphics[width=0.45\textwidth]{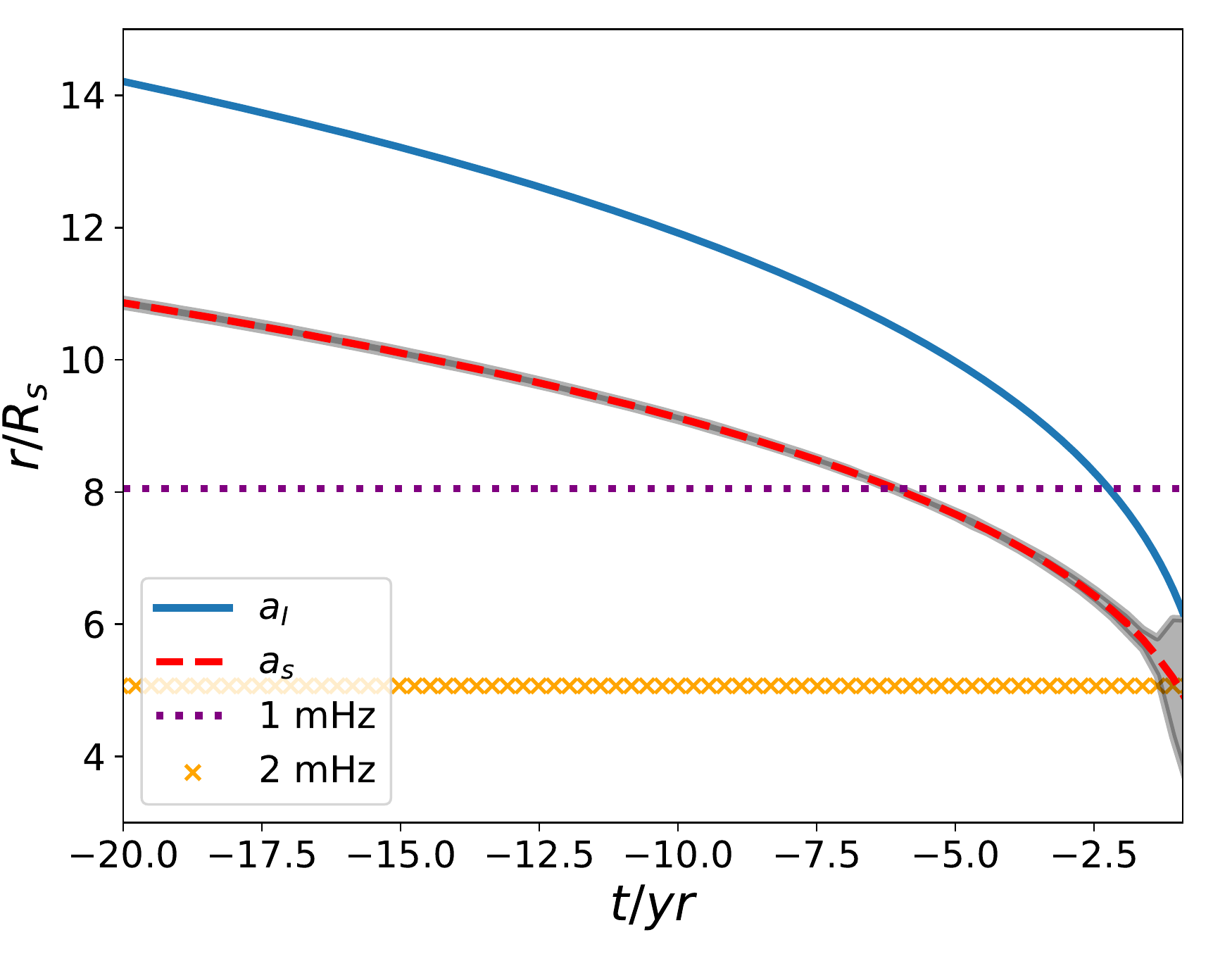}
\caption{Evolution of an EMRI-IMRI pair in the GW-radiation regime.  The $x$
axis shows the time before the merger of the IMBH and the SMBH.  The blue solid
(red dashed) line refers to the semimajor axis of the IMBH (sBH).  The gray shaded
area shows the radial range reachable by the sBH, which is determined by the
orbital apocenter and pericenter.  The purple dotted line and the
yellow crosses mark the radii where the fundamental frequency of the GW is,
respectively,  $1$ and $2$ mHz. }
    \label{fig:EMRI-IMRI_1}
\end{figure}

\begin{figure}
    \centering
    \includegraphics[width=0.45\textwidth]{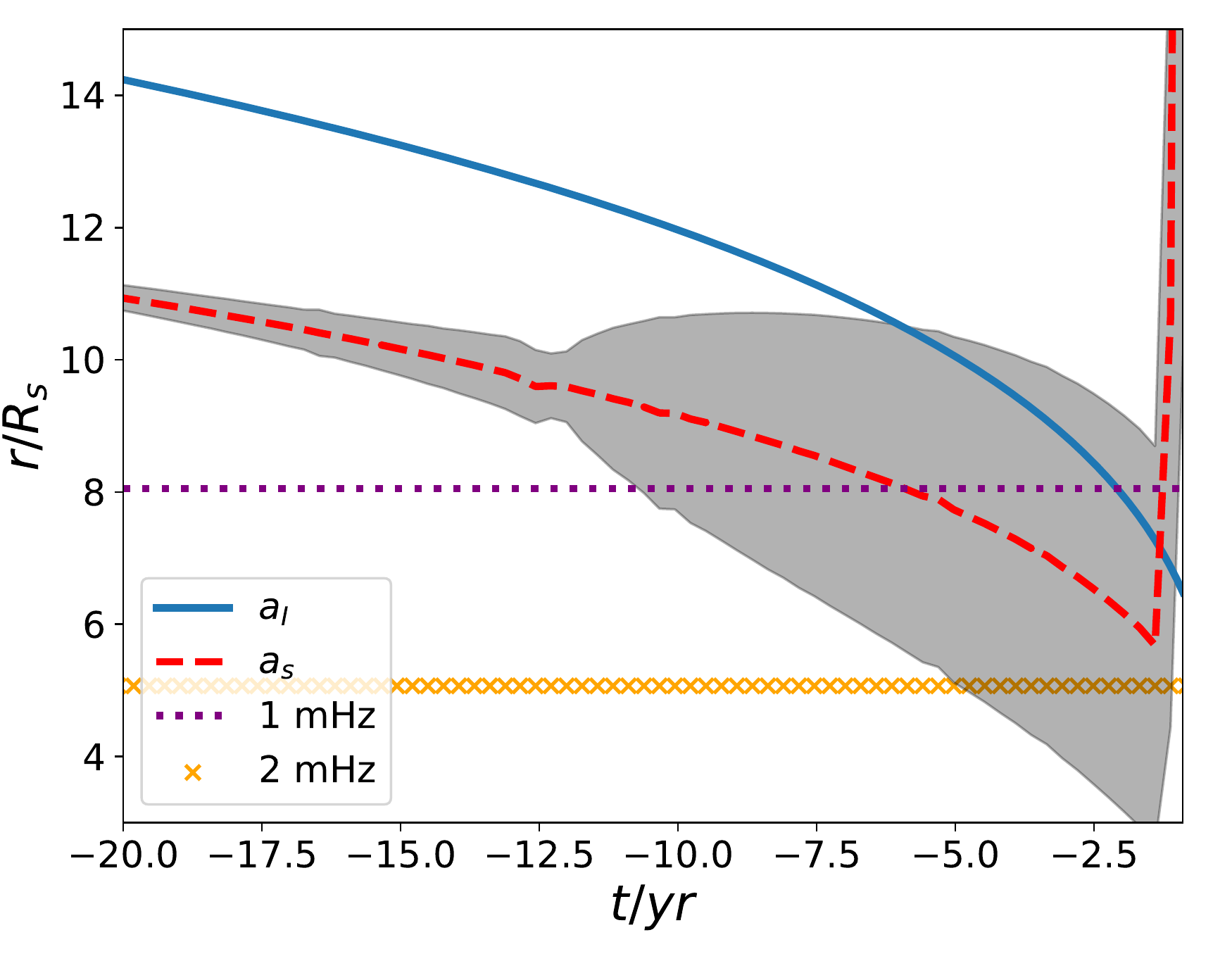}
\caption{The same as Figure~\ref{fig:EMRI-IMRI_2}, but the damping timescale of
the orbital eccentricity of the sBH is three times longer than that in
Figure~\ref{fig:EMRI-IMRI_1}. } \label{fig:EMRI-IMRI_2} 
\end{figure}

In our 2D simulation, the initial loci of the sBH and the IMBH are $a_s= 23
R_{\rm{S}}$ and $a_I = 30 R_{\rm{S}}$, which are adopted from our 1D
simulation. In this configuration, the two BHs are in the 3:2 resonance. The
initial $e_s$ is chosen as $0.01$, to be consistent with the output of the 1D
simulation. The exact value does not significantly affect the result.

Figure~\ref{fig:EMRI-IMRI_1} shows the evolution of the IMBH (blue solid line)
and the inner sBH (red dashed line) in the 2D simulation with our default
damping timescale.  We find that the migration of the two BHs are still
synchronized.  More importantly, the sBH and the IMBH successively enters the
LISA band (below the purple dotted line) with a time delay of about four years.
This result indicates that within its mission duration of about five years
\citep{amaro17}, LISA may detect an EMRI and an IMRI from the same AGN.

About one year after the IMBH enters the LISA band, we find a fast excitation
of the orbital eccentricity of the sBH. This is caused by an imbalance between
the increasingly strong tidal torque from the IMBH and the insufficient damping
rate of the gaseous disk.  We see a final break up of the EMRI-IMRI pair by the
end of our simulation (also see Figure~\ref{fig:EMRI-IMRI_2}). Therefore, we
conclude that the EMRI and IMRI could simultaneously appear in the LISA band
for about one year.  After that, the EMRI signal will disappear but the IMBH
will continue to merge with the central SMBH. In our fiducial model, the sBH,
because of GW radiation, will reenter the LISA band thousands of years after
the merger of the IMBH with the SMBH.

Figure \ref{fig:EMRI-IMRI_2} shows the result with a damping timescale (for
$e_s$) three times longer than the default one. We can see that now the
eccentricity of the sBH is excited at a much earlier stage.  The high
eccentricity allows the sBH to enter the LISA band earlier, because (i) the
highest GW frequency is determined by the pericenter of the orbit
\citep{peters63} and (ii) the pericenter becomes sufficiently small at an
earlier time.  In fact, in our simulation we find that the pericenter of the
sBH already enters the sensitive band of LISA ($>1$ mHz) eight years before the
IMBH does, and $10$ years before the final coalescence between the IMBH and the
SMBH.  Therefore, LISA can also simultaneously detect an EMRI and an IMRI in
this case.  Similar to what we have found in the previous case with a short
damping timescale, the sBH is eventually scattered to a much larger radius by
the IMBH. Reentry of the sBH into the LISA happens more than 4000 years later.
Therefore, LISA could not detect both this event and the earlier IMRI event.

\section{Discussion and conclusion \label{sec:conclusion}}

In this work, we studied the interaction between a gap-opening IMBH in an AGN
disk and the sBHs embedded in the disk. We found that during its Type-II
migration the IMBH could capture a sBH inside its orbit and form an EMRI-IMRI
pair.  Our simulations showed that the subsequent migration of the two BHs is
synchronized until they reach a distance of about $\sim$ 10 $R_{\rm{S}}$ from
the central SMBH. Consequently, the EMRI and the IMRI successively enter the
LISA band within a timescale of 4-8 years. The time delay depends on the
effectiveness of the hydrodynamical damping of the orbital eccentricity of the
sBH.  When the damping effect is strong, the EMRI and the IMRI could even
simultaneously appear in the LISA band for about one year before the EMRI
is disrupted by the gravitational force of the IMBH.

Estimating the event rate of such EMRI-IMRI pairs is difficult given many
theoretical uncertainties, such as the number of sBHs inside an AGN accretion
disk and the formation channel of the IMBHs.  If the IMBHs are produced by
successive mergers of the sBHs inside migration traps, we can estimate the
event rate in the following way. Recent theoretical models neglecting migration
traps predict that the event rate of the EMRIs in AGNs is $10$--$10^4$ per year
\citep{pan21b}.  If a migration trap is present in the disk, the sBHs will no
longer form EMRIs but accumulate in the trap. Since on average $10$ sBHs are
needed to form a gap-opening IMBH which can leave the migration trap
(Section~\ref{sec:single}), the formation rate of the EMRI-IMRI pairs would be
about $1-10^3$ per year in this scenario.

Given the mutual interaction between the EMRI and the IMRI in the pair, we
expect the GW signal to be different from either a single EMRI or a single
IMRI.  Some hints can be drawn from the previous studies on the perturbation of
an EMRI by the nearby stars or sBHs.  \cite{2012ApJ...744L..20A, 2012PhRvL.109g1102F,
2019PhRvL.123j1103B, 2021PhRvD.104d4056G} showed that the effect is the
strongest when the perturber and the EMRI temporarily enter a resonance, which
will result in a significant phase shift in the waveform of the EMRI.
\cite{Gupta22} further showed that such a phase shift, if ignored in the
waveform model, will induce non-negligible biases in the estimated parameters
of the EMRI.  However, if properly accounted for, the perturbed signal can
reveal the mass and orbital parameters of the perturber \citep{Speri21,
Gupta22}.  These earlier results also apply to our EMRI-IMRI pairs since they
are locked in resonances for an extended period of time, as we have seen in
Section~\ref{subsec:later}.  Moreover, because the EMRIs enter the LISA band
earlier than the IMRIs, the EMRI signals could reveal the outer IMBHs  and be
used to infer their parameters even before the IMRIs enter the LISA band.

Although we only run simulations with $m_{\rm{IMBH}} = 100 M_{\odot}$, the
results are applicable to more massive IMBHs as long as $m_{\rm{IMBH}}$ is not
much larger than the local disk mass $M_d\equiv 4 \pi \Sigma a_{I}^2$.  When
$m_{\rm{IMBH}}\gg M_d$, the IMBH cannot follow the viscous evolution of the
disk because the gas can no longer efficiently absorb the angular momentum of
the IMBH \citep{Syer95}.  In this case the migration timescale of the IMBH will
be much longer than the $T_{II}$ in Equation~(\ref{eq:T_II}) and the Type-I
migration timescale in Equation~(\ref{eq:T_I}).  Therefore, the IMBH cannot
catch up with the sBHs migrating inside its orbit. The pair of an outer IMBH
and an inner EMRI will not form in this case.

However, when $m_{\rm{IMBH}}\gg M_d$, the slowly migrating IMBH, by opening a
gap in the disk, would block the fast migrating sBHs outside the orbit of the
IMBH.  Therefore, a new type of trap forms, at the outer boundary of the gap.
We did not study this type of trap in this work because it disappears as soon
as the inner IMBH enters the GW-radiation region, when the migration of the
IMBH becomes faster than the Type-I migration of the sBHs (e.g., see the green
dotted line in Figure~\ref{fig:migration}). Nevertheless, before the trap
disappears, sBHs will accumulate inside it, interact with each other, and form
binaries, just as they will do in a conventional Type-I migration trap
\citep{secunda19}. Therefore, we expect a positive correlation between the
formation of SMBH-IMBH binaries and an enhancement of the merger rate of sBHs
in AGNs.  It is worth noting that  such a trap exists not only around a massive
IMBH, but also around a secondary SMBH if there is one inside an AGN disk.

In conclusion, we find that gap-opening IMBHs (or secondary SMBHs) in AGN disks
could trap sBHs around their orbits. The traps could produce EMRI-IMRI pairs
which may appear simultaneously in the LISA band.  Detecting such pairs
can shed light on the formation and evolution of the compact objects in AGN
disk. A successful detection also requires more efforts in modeling the
three-body dynamics and the corresponding GW signals in the region of strong
gravity.

\section*{Acknowledgements}

This work is supported by  the National Key Research and Development Program of
China Grant No.  2021YFC2203002, and the National Science Foundation of China
grants No. 11991053 and 11873022. The computation in this work was performed on
the High Performance Computing Platform of the Centre for Life Science, Peking
University. The authors would like to thank Camilo Fontecilla and Jorge Cuadra
for sharing their code.

\bibliographystyle{aasjournal}
\bibliography{bibbase,mybib,References}{}

\begin{thebibliography}{}
\expandafter\ifx\csname natexlab\endcsname\relax\def\natexlab#1{#1}\fi
\providecommand{\url}[1]{\href{#1}{#1}}
\providecommand{\dodoi}[1]{doi:~\href{http://doi.org/#1}{\nolinkurl{#1}}}
\providecommand{\doeprint}[1]{\href{http://ascl.net/#1}{\nolinkurl{http://ascl.net/#1}}}
\providecommand{\doarXiv}[1]{\href{https://arxiv.org/abs/#1}{\nolinkurl{https://arxiv.org/abs/#1}}}

\bibitem[{{Amaro-Seoane}(2018)}]{pau18}
{Amaro-Seoane}, P. 2018, Living Reviews in Relativity, 21, 4,
  \dodoi{10.1007/s41114-018-0013-8}

\bibitem[{{Amaro-Seoane} {et~al.}(2012){Amaro-Seoane}, {Brem}, {Cuadra}, \&
  {Armitage}}]{2012ApJ...744L..20A}
{Amaro-Seoane}, P., {Brem}, P., {Cuadra}, J., \& {Armitage}, P.~J. 2012, \apjl,
  744, L20, \dodoi{10.1088/2041-8205/744/2/L20}

\bibitem[{{Amaro-Seoane} {et~al.}(2007){Amaro-Seoane}, {Gair}, {Freitag},
  {Miller}, {Mandel}, {Cutler}, \& {Babak}}]{2007CQGra..24R.113A}
{Amaro-Seoane}, P., {Gair}, J.~R., {Freitag}, M., {et~al.} 2007, Classical and
  Quantum Gravity, 24, R113, \dodoi{10.1088/0264-9381/24/17/R01}

\bibitem[{{Amaro-Seoane} {et~al.}(2017){Amaro-Seoane}, {Audley}, {Babak},
  {Baker}, {Barausse}, {Bender}, {Berti}, {Binetruy}, {Born}, \&
  {Bortoluzzi}}]{amaro17}
{Amaro-Seoane}, P., {Audley}, H., {Babak}, S., {et~al.} 2017, ArXiv e-prints.
\newblock \doarXiv{1702.00786}

\bibitem[{{Antoni} {et~al.}(2019){Antoni}, {MacLeod}, \&
  {Ramirez-Ruiz}}]{antoni19}
{Antoni}, A., {MacLeod}, M., \& {Ramirez-Ruiz}, E. 2019, \apj, 884, 22,
  \dodoi{10.3847/1538-4357/ab3466}

\bibitem[{{Armitage} \& {Natarajan}(2002)}]{armitage02}
{Armitage}, P.~J., \& {Natarajan}, P. 2002, \apjl, 567, L9,
  \dodoi{10.1086/339770}

\bibitem[{{Artymowicz} {et~al.}(1993){Artymowicz}, {Lin}, \&
  {Wampler}}]{artymowicz93}
{Artymowicz}, P., {Lin}, D.~N.~C., \& {Wampler}, E.~J. 1993, \apj, 409, 592,
  \dodoi{10.1086/172690}

\bibitem[{{Artymowicz} \& {Lubow}(1994)}]{artymowicz94}
{Artymowicz}, P., \& {Lubow}, S.~H. 1994, \apj, 421, 651,
  \dodoi{10.1086/173679}

\bibitem[{{Bartos} {et~al.}(2017){Bartos}, {Kocsis}, {Haiman}, \&
  {M{\'a}rka}}]{bartos17}
{Bartos}, I., {Kocsis}, B., {Haiman}, Z., \& {M{\'a}rka}, S. 2017, \apj, 835,
  165, \dodoi{10.3847/1538-4357/835/2/165}

\bibitem[{{Baruteau} {et~al.}(2011){Baruteau}, {Cuadra}, \& {Lin}}]{baruteau11}
{Baruteau}, C., {Cuadra}, J., \& {Lin}, D.~N.~C. 2011, \apj, 726, 28,
  \dodoi{10.1088/0004-637X/726/1/28}

\bibitem[{{Basu} {et~al.}(2008){Basu}, {Mondal}, \& {Chakrabarti}}]{basu08}
{Basu}, P., {Mondal}, S., \& {Chakrabarti}, S.~K. 2008, \mnras, 388, 219,
  \dodoi{10.1111/j.1365-2966.2008.13368.x}

\bibitem[{{Bellovary} {et~al.}(2016){Bellovary}, {Mac Low}, {McKernan}, \&
  {Ford}}]{bellovary16}
{Bellovary}, J.~M., {Mac Low}, M.-M., {McKernan}, B., \& {Ford}, K.~E.~S. 2016,
  \apjl, 819, L17, \dodoi{10.3847/2041-8205/819/2/L17}

\bibitem[{{Bitsch} \& {Kley}(2010)}]{2010A&A...523A..30B}
{Bitsch}, B., \& {Kley}, W. 2010, \aap, 523, A30,
  \dodoi{10.1051/0004-6361/201014414}

\bibitem[{{Bonga} {et~al.}(2019){Bonga}, {Yang}, \&
  {Hughes}}]{2019PhRvL.123j1103B}
{Bonga}, B., {Yang}, H., \& {Hughes}, S.~A. 2019, \prl, 123, 101103,
  \dodoi{10.1103/PhysRevLett.123.101103}

\bibitem[{{Chakrabarti}(1993)}]{chakrabarti93}
{Chakrabarti}, S.~K. 1993, \apj, 411, 610, \dodoi{10.1086/172863}

\bibitem[{{Coleman} \& {Nelson}(2014)}]{2014MNRAS.445..479C}
{Coleman}, G. A.~L., \& {Nelson}, R.~P. 2014, \mnras, 445, 479,
  \dodoi{10.1093/mnras/stu1715}

\bibitem[{{Cresswell} {et~al.}(2007){Cresswell}, {Dirksen}, {Kley}, \&
  {Nelson}}]{2007A&A...473..329C}
{Cresswell}, P., {Dirksen}, G., {Kley}, W., \& {Nelson}, R.~P. 2007, \aap, 473,
  329, \dodoi{10.1051/0004-6361:20077666}

\bibitem[{{Cresswell} \& {Nelson}(2008)}]{2008A&A...482..677C}
{Cresswell}, P., \& {Nelson}, R.~P. 2008, \aap, 482, 677,
  \dodoi{10.1051/0004-6361:20079178}

\bibitem[{{Derdzinski} \& {Mayer}(2022)}]{2022arXiv220510382D}
{Derdzinski}, A., \& {Mayer}, L. 2022, arXiv e-prints, arXiv:2205.10382.
\newblock \doarXiv{2205.10382}

\bibitem[{{Fabj} {et~al.}(2020){Fabj}, {Nasim}, {Caban}, {Ford}, {McKernan}, \&
  {Bellovary}}]{2020MNRAS.499.2608F}
{Fabj}, G., {Nasim}, S.~S., {Caban}, F., {et~al.} 2020, \mnras, 499, 2608,
  \dodoi{10.1093/mnras/staa3004}

\bibitem[{{Flanagan} \& {Hinderer}(2012)}]{2012PhRvL.109g1102F}
{Flanagan}, {\'E}.~{\'E}., \& {Hinderer}, T. 2012, \prl, 109, 071102,
  \dodoi{10.1103/PhysRevLett.109.071102}

\bibitem[{{Fontecilla} {et~al.}(2019){Fontecilla}, {Haiman}, \&
  {Cuadra}}]{Fontecilla19}
{Fontecilla}, C., {Haiman}, Z., \& {Cuadra}, J. 2019, mnras, 482, 4383,
  \dodoi{10.1093/mnras/sty2972}

\bibitem[{{Ford} {et~al.}(2000){Ford}, {Kozinsky}, \&
  {Rasio}}]{2000ApJ...535..385F}
{Ford}, E.~B., {Kozinsky}, B., \& {Rasio}, F.~A. 2000, \apj, 535, 385,
  \dodoi{10.1086/308815}

\bibitem[{{Gabor} \& {Bournaud}(2013)}]{2013MNRAS.434..606G}
{Gabor}, J.~M., \& {Bournaud}, F. 2013, \mnras, 434, 606,
  \dodoi{10.1093/mnras/stt1046}

\bibitem[{{Gayathri} {et~al.}(2020){Gayathri}, {Bartos}, {Haiman}, {Klimenko},
  {Kocsis}, {M{\'a}rka}, \& {Yang}}]{gayathri20}
{Gayathri}, V., {Bartos}, I., {Haiman}, Z., {et~al.} 2020, \apjl, 890, L20,
  \dodoi{10.3847/2041-8213/ab745d}

\bibitem[{{Gerosa} \& {Fishbach}(2021)}]{2021NatAs...5..749G}
{Gerosa}, D., \& {Fishbach}, M. 2021, Nature Astronomy, 5, 749,
  \dodoi{10.1038/s41550-021-01398-w}

\bibitem[{{Goldreich} \& {Tremaine}(1978)}]{1978ApJ...222..850G}
{Goldreich}, P., \& {Tremaine}, S. 1978, \apj, 222, 850, \dodoi{10.1086/156203}

\bibitem[{{Goldreich} \& {Tremaine}(1979)}]{1979ApJ...233..857G}
---. 1979, \apj, 233, 857, \dodoi{10.1086/157448}

\bibitem[{{Goldreich} \& {Tremaine}(1980)}]{1980ApJ...241..425G}
---. 1980, \apj, 241, 425, \dodoi{10.1086/158356}

\bibitem[{{Gon{\c{c}}alves} {et~al.}(2008){Gon{\c{c}}alves}, {Steidel}, \&
  {Pettini}}]{2008ApJ...676..816G}
{Gon{\c{c}}alves}, T.~S., {Steidel}, C.~C., \& {Pettini}, M. 2008, \apj, 676,
  816, \dodoi{10.1086/527313}

\bibitem[{{Goodman}(2003)}]{Goodman2003}
{Goodman}, J. 2003, \mnras, 339, 937, \dodoi{10.1046/j.1365-8711.2003.06241.x}

\bibitem[{{Goodman} \& {Tan}(2004)}]{Goodman2004}
{Goodman}, J., \& {Tan}, J.~C. 2004, \apj, 608, 108, \dodoi{10.1086/386360}

\bibitem[{{Gould} \& {Rix}(2000)}]{gould00}
{Gould}, A., \& {Rix}, H.-W. 2000, \apjl, 532, L29, \dodoi{10.1086/312562}

\bibitem[{{Gr{\"o}bner} {et~al.}(2020){Gr{\"o}bner}, {Ishibashi}, {Tiwari},
  {Haney}, \& {Jetzer}}]{grobner20}
{Gr{\"o}bner}, M., {Ishibashi}, W., {Tiwari}, S., {Haney}, M., \& {Jetzer}, P.
  2020, \aap, 638, A119, \dodoi{10.1051/0004-6361/202037681}

\bibitem[{{Gupta} {et~al.}(2021){Gupta}, {Bonga}, {Chua}, \&
  {Tanaka}}]{2021PhRvD.104d4056G}
{Gupta}, P., {Bonga}, B., {Chua}, A. J.~K., \& {Tanaka}, T. 2021, \prd, 104,
  044056, \dodoi{10.1103/PhysRevD.104.044056}

\bibitem[{{Gupta} {et~al.}(2022){Gupta}, {Speri}, {Bonga}, {Chua}, \&
  {Tanaka}}]{Gupta22}
{Gupta}, P., {Speri}, L., {Bonga}, B., {Chua}, A. J.~K., \& {Tanaka}, T. 2022,
  arXiv e-prints, arXiv:2205.04808.
\newblock \doarXiv{2205.04808}

\bibitem[{{Hahn} \& {Ward}(1996)}]{1996LPI....27..479H}
{Hahn}, J.~M., \& {Ward}, W.~R. 1996, in Lunar and Planetary Science
  Conference, Vol.~27, Lunar and Planetary Science Conference, 479

\bibitem[{{Hopkins} \& {Hernquist}(2009)}]{2009ApJ...698.1550H}
{Hopkins}, P.~F., \& {Hernquist}, L. 2009, \apj, 698, 1550,
  \dodoi{10.1088/0004-637X/698/2/1550}

\bibitem[{{Ida} {et~al.}(2020){Ida}, {Muto}, {Matsumura}, \&
  {Brasser}}]{2020MNRAS.494.5666I}
{Ida}, S., {Muto}, T., {Matsumura}, S., \& {Brasser}, R. 2020, \mnras, 494,
  5666, \dodoi{10.1093/mnras/staa1073}

\bibitem[{{Ivanov} {et~al.}(1999{\natexlab{a}}){Ivanov}, {Papaloizou}, \&
  {Polnarev}}]{Ivanov1999}
{Ivanov}, P.~B., {Papaloizou}, J.~C.~B., \& {Polnarev}, A.~G.
  1999{\natexlab{a}}, \mnras, 307, 79, \dodoi{10.1046/j.1365-8711.1999.02623.x}

\bibitem[{{Ivanov} {et~al.}(1999{\natexlab{b}}){Ivanov}, {Papaloizou}, \&
  {Polnarev}}]{ivanov99}
---. 1999{\natexlab{b}}, \mnras, 307, 79,
  \dodoi{10.1046/j.1365-8711.1999.02623.x}

\bibitem[{{Karas} \& {{\v{S}}ubr}(2001)}]{karas01}
{Karas}, V., \& {{\v{S}}ubr}, L. 2001, \aap, 376, 686,
  \dodoi{10.1051/0004-6361:20011009}

\bibitem[{{Kocsis} {et~al.}(2012){Kocsis}, {Haiman}, \& {Loeb}}]{kocsis12}
{Kocsis}, B., {Haiman}, Z., \& {Loeb}, A. 2012, \mnras, 427, 2680,
  \dodoi{10.1111/j.1365-2966.2012.22118.x}

\bibitem[{{Kocsis} {et~al.}(2011){Kocsis}, {Yunes}, \& {Loeb}}]{kocsis11}
{Kocsis}, B., {Yunes}, N., \& {Loeb}, A. 2011, \prd, 84, 024032,
  \dodoi{10.1103/PhysRevD.84.024032}

\bibitem[{{Leigh} {et~al.}(2018){Leigh}, {Geller}, {McKernan}, {Ford}, {Mac
  Low}, {Bellovary}, {Haiman}, {Lyra}, {Samsing}, {O'Dowd}, {Kocsis}, \&
  {Endlich}}]{leigh18}
{Leigh}, N.~W.~C., {Geller}, A.~M., {McKernan}, B., {et~al.} 2018, \mnras, 474,
  5672, \dodoi{10.1093/mnras/stx3134}

\bibitem[{{Levin}(2003)}]{levin03}
{Levin}, Y. 2003, arXiv e-prints, astro.
\newblock \doarXiv{astro-ph/0307084}

\bibitem[{{Lin} \& {Papaloizou}(1979{\natexlab{a}})}]{1979MNRAS.186..799L}
{Lin}, D.~N.~C., \& {Papaloizou}, J. 1979{\natexlab{a}}, mnras, 186, 799,
  \dodoi{10.1093/mnras/186.4.799}

\bibitem[{{Lin} \& {Papaloizou}(1979{\natexlab{b}})}]{1979MNRAS.188..191L}
---. 1979{\natexlab{b}}, mnras, 188, 191, \dodoi{10.1093/mnras/188.2.191}

\bibitem[{{Lin} \& {Papaloizou}(1986{\natexlab{a}})}]{lin86}
---. 1986{\natexlab{a}}, \apj, 309, 846, \dodoi{10.1086/164653}

\bibitem[{{Lin} \& {Papaloizou}(1986{\natexlab{b}})}]{1986ApJ...309..846L}
---. 1986{\natexlab{b}}, \apj, 309, 846, \dodoi{10.1086/164653}

\bibitem[{{Mastrobuono-Battisti} {et~al.}(2014){Mastrobuono-Battisti},
  {Perets}, \& {Loeb}}]{2014ApJ...796...40M}
{Mastrobuono-Battisti}, A., {Perets}, H.~B., \& {Loeb}, A. 2014, \apj, 796, 40,
  \dodoi{10.1088/0004-637X/796/1/40}

\bibitem[{{McKernan} {et~al.}(2012{\natexlab{a}}){McKernan}, {Ford}, {Lyra}, \&
  {Perets}}]{2012MNRAS.425..460M}
{McKernan}, B., {Ford}, K.~E.~S., {Lyra}, W., \& {Perets}, H.~B.
  2012{\natexlab{a}}, \mnras, 425, 460,
  \dodoi{10.1111/j.1365-2966.2012.21486.x}

\bibitem[{{McKernan} {et~al.}(2012{\natexlab{b}}){McKernan}, {Ford}, {Lyra}, \&
  {Perets}}]{mckernan12}
---. 2012{\natexlab{b}}, \mnras, 425, 460,
  \dodoi{10.1111/j.1365-2966.2012.21486.x}

\bibitem[{{McKernan} {et~al.}(2011){McKernan}, {Ford}, {Lyra}, {Perets},
  {Winter}, \& {Yaqoob}}]{mckernan11}
{McKernan}, B., {Ford}, K.~E.~S., {Lyra}, W., {et~al.} 2011, \mnras, 417, L103,
  \dodoi{10.1111/j.1745-3933.2011.01132.x}

\bibitem[{McKernan {et~al.}(2018)McKernan, Ford, Bellovary, Leigh, Haiman,
  Kocsis, Lyra, Low, Metzger, O'Dowd, Endlich, \& Rosen}]{mckernan18}
McKernan, B., Ford, K. E.~S., Bellovary, J., {et~al.} 2018, The Astrophysical
  Journal, 866, 66, \dodoi{10.3847/1538-4357/aadae5}

\bibitem[{{Miller} \& {Lauburg}(2009)}]{miller09a}
{Miller}, M.~C., \& {Lauburg}, V.~M. 2009, apj, 692, 917,
  \dodoi{10.1088/0004-637X/692/1/917}

\bibitem[{{Miranda} {et~al.}(2017){Miranda}, {Mu{\~n}oz}, \&
  {Lai}}]{2017MNRAS.466.1170M}
{Miranda}, R., {Mu{\~n}oz}, D.~J., \& {Lai}, D. 2017, \mnras, 466, 1170,
  \dodoi{10.1093/mnras/stw3189}

\bibitem[{{Molteni} {et~al.}(1994){Molteni}, {Gerardi}, \&
  {Chakrabarti}}]{molteni94}
{Molteni}, D., {Gerardi}, G., \& {Chakrabarti}, S.~K. 1994, \apj, 436, 249,
  \dodoi{10.1086/174897}

\bibitem[{{Mu{\~n}oz} {et~al.}(2020){Mu{\~n}oz}, {Lai}, {Kratter}, \&
  {Miranda}}]{2020ApJ...889..114M}
{Mu{\~n}oz}, D.~J., {Lai}, D., {Kratter}, K., \& {Miranda}, R. 2020, \apj, 889,
  114, \dodoi{10.3847/1538-4357/ab5d33}

\bibitem[{{Murray} \& {Dermott}(1999)}]{1999ssd..book.....M}
{Murray}, C.~D., \& {Dermott}, S.~F. 1999, {Solar system dynamics}

\bibitem[{{Muto} {et~al.}(2011){Muto}, {Takeuchi}, \&
  {Ida}}]{2011ApJ...737...37M}
{Muto}, T., {Takeuchi}, T., \& {Ida}, S. 2011, \apj, 737, 37,
  \dodoi{10.1088/0004-637X/737/1/37}

\bibitem[{{Nasim} {et~al.}(2022){Nasim}, {Fabj}, {Caban}, {Secunda}, {Ford},
  {McKernan}, {Bellovary}, {Leigh}, \& {Lyra}}]{2022arXiv220709540N}
{Nasim}, S.~S., {Fabj}, G., {Caban}, F., {et~al.} 2022, arXiv e-prints,
  arXiv:2207.09540.
\newblock \doarXiv{2207.09540}

\bibitem[{{Paardekooper} {et~al.}(2010){Paardekooper}, {Baruteau}, {Crida}, \&
  {Kley}}]{2010MNRAS.401.1950P}
{Paardekooper}, S.~J., {Baruteau}, C., {Crida}, A., \& {Kley}, W. 2010, \mnras,
  401, 1950, \dodoi{10.1111/j.1365-2966.2009.15782.x}

\bibitem[{{Paardekooper} {et~al.}(2011){Paardekooper}, {Baruteau}, \&
  {Kley}}]{Paardekooper11}
{Paardekooper}, S.~J., {Baruteau}, C., \& {Kley}, W. 2011, \mnras, 410, 293,
  \dodoi{10.1111/j.1365-2966.2010.17442.x}

\bibitem[{{Pan} {et~al.}(2021){Pan}, {Lyu}, \& {Yang}}]{pan21b}
{Pan}, Z., {Lyu}, Z., \& {Yang}, H. 2021, arXiv e-prints, arXiv:2104.01208.
\newblock \doarXiv{2104.01208}

\bibitem[{{Pan} \& {Yang}(2021)}]{pan21}
{Pan}, Z., \& {Yang}, H. 2021, \prd, 103, 103018,
  \dodoi{10.1103/PhysRevD.103.103018}

\bibitem[{{Papaloizou} \& {Larwood}(2000)}]{2000MNRAS.315..823P}
{Papaloizou}, J.~C.~B., \& {Larwood}, J.~D. 2000, \mnras, 315, 823,
  \dodoi{10.1046/j.1365-8711.2000.03466.x}

\bibitem[{{Peng} \& {Chen}(2021)}]{peng21}
{Peng}, P., \& {Chen}, X. 2021, \mnras, 505, 1324,
  \dodoi{10.1093/mnras/stab1419}

\bibitem[{{Peters} \& {Mathews}(1963)}]{peters63}
{Peters}, P.~C., \& {Mathews}, J. 1963, Physical Review, 131, 435,
  \dodoi{10.1103/PhysRev.131.435}

\bibitem[{{Pierens} \& {Nelson}(2008)}]{2008A&A...482..333P}
{Pierens}, A., \& {Nelson}, R.~P. 2008, aap, 482, 333,
  \dodoi{10.1051/0004-6361:20079062}

\bibitem[{{Podlewska} \& {Szuszkiewicz}(2008)}]{2008MNRAS.386.1347P}
{Podlewska}, E., \& {Szuszkiewicz}, E. 2008, \mnras, 386, 1347,
  \dodoi{10.1111/j.1365-2966.2008.12871.x}

\bibitem[{{Rein} \& {Liu}(2012)}]{rebound}
{Rein}, H., \& {Liu}, S.~F. 2012, \aap, 537, A128,
  \dodoi{10.1051/0004-6361/201118085}

\bibitem[{{Rein} \& {Spiegel}(2015)}]{reboundias15}
{Rein}, H., \& {Spiegel}, D.~S. 2015, \mnras, 446, 1424,
  \dodoi{10.1093/mnras/stu2164}

\bibitem[{{S{\'a}nchez-Salcedo}(2020)}]{sanchez20}
{S{\'a}nchez-Salcedo}, F.~J. 2020, \apj, 897, 142,
  \dodoi{10.3847/1538-4357/ab9b2d}

\bibitem[{{Secunda} {et~al.}(2019){Secunda}, {Bellovary}, {Mac Low}, {Ford},
  {McKernan}, {Leigh}, {Lyra}, \& {S{\'a}ndor}}]{secunda19}
{Secunda}, A., {Bellovary}, J., {Mac Low}, M.-M., {et~al.} 2019, \apj, 878, 85,
  \dodoi{10.3847/1538-4357/ab20ca}

\bibitem[{{Secunda} {et~al.}(2020){Secunda}, {Bellovary}, {Mac Low}, {Ford},
  {McKernan}, {Leigh}, {Lyra}, {S{\'a}ndor}, \& {Adorno}}]{secunda20}
---. 2020, \apj, 903, 133, \dodoi{10.3847/1538-4357/abbc1d}

\bibitem[{{Shlosman} \& {Begelman}(1987)}]{1987Natur.329..810S}
{Shlosman}, I., \& {Begelman}, M.~C. 1987, \nat, 329, 810,
  \dodoi{10.1038/329810a0}

\bibitem[{{Sigl} {et~al.}(2007){Sigl}, {Schnittman}, \& {Buonanno}}]{sigl07}
{Sigl}, G., {Schnittman}, J., \& {Buonanno}, A. 2007, \prd, 75, 024034,
  \dodoi{10.1103/PhysRevD.75.024034}

\bibitem[{{Speri} \& {Gair}(2021)}]{Speri21}
{Speri}, L., \& {Gair}, J.~R. 2021, \prd, 103, 124032,
  \dodoi{10.1103/PhysRevD.103.124032}

\bibitem[{{Stone} {et~al.}(2017){Stone}, {Metzger}, \& {Haiman}}]{stone17}
{Stone}, N.~C., {Metzger}, B.~D., \& {Haiman}, Z. 2017, \mnras, 464, 946,
  \dodoi{10.1093/mnras/stw2260}

\bibitem[{{Syer} \& {Clarke}(1995)}]{Syer95}
{Syer}, D., \& {Clarke}, C.~J. 1995, \mnras, 277, 758,
  \dodoi{10.1093/mnras/277.3.758}

\bibitem[{{Syer} {et~al.}(1991){Syer}, {Clarke}, \& {Rees}}]{syer91}
{Syer}, D., {Clarke}, C.~J., \& {Rees}, M.~J. 1991, \mnras, 250, 505,
  \dodoi{10.1093/mnras/250.3.505}

\bibitem[{{Tagawa} {et~al.}(2021){Tagawa}, {Haiman}, {Bartos}, {Kocsis}, \&
  {Omukai}}]{2021MNRAS.507.3362T}
{Tagawa}, H., {Haiman}, Z., {Bartos}, I., {Kocsis}, B., \& {Omukai}, K. 2021,
  \mnras, 507, 3362, \dodoi{10.1093/mnras/stab2315}

\bibitem[{{Tagawa} {et~al.}(2020){Tagawa}, {Haiman}, \& {Kocsis}}]{tagawa20}
{Tagawa}, H., {Haiman}, Z., \& {Kocsis}, B. 2020, \apj, 898, 25,
  \dodoi{10.3847/1538-4357/ab9b8c}

\bibitem[{{Tanaka} \& {Ward}(2004)}]{2004ApJ...602..388T}
{Tanaka}, H., \& {Ward}, W.~R. 2004, apj, 602, 388, \dodoi{10.1086/380992}

\bibitem[{{Volonteri} {et~al.}(2003){Volonteri}, {Haardt}, \&
  {Madau}}]{volonteri03}
{Volonteri}, M., {Haardt}, F., \& {Madau}, P. 2003, \apj, 582, 559,
  \dodoi{10.1086/344675}

\bibitem[{{{\v{S}}ubr} \& {Karas}(1999)}]{subr99}
{{\v{S}}ubr}, L., \& {Karas}, V. 1999, \aap, 352, 452.
\newblock \doarXiv{astro-ph/9910401}

\bibitem[{{Ward}(1997)}]{1997Icar..126..261W}
{Ward}, W.~R. 1997, \icarus, 126, 261, \dodoi{10.1006/icar.1996.5647}

\bibitem[{{Yang} {et~al.}(2019{\natexlab{a}}){Yang}, {Bonga}, {Peng}, \&
  {Li}}]{Yang2019}
{Yang}, H., {Bonga}, B., {Peng}, Z., \& {Li}, G. 2019{\natexlab{a}}, \prd, 100,
  124056, \dodoi{10.1103/PhysRevD.100.124056}

\bibitem[{{Yang} {et~al.}(2019{\natexlab{b}}){Yang}, {Bartos}, {Gayathri},
  {Ford}, {Haiman}, {Klimenko}, {Kocsis}, {M{\'a}rka}, {M{\'a}rka}, {McKernan},
  \& {O'Shaughnessy}}]{yang19}
{Yang}, Y., {Bartos}, I., {Gayathri}, V., {et~al.} 2019{\natexlab{b}}, \prl,
  123, 181101, \dodoi{10.1103/PhysRevLett.123.181101}

\bibitem[{{Yang} {et~al.}(2019{\natexlab{c}}){Yang}, {Bartos}, {Gayathri},
  {Ford}, {Haiman}, {Klimenko}, {Kocsis}, {M{\'a}rka}, {M{\'a}rka}, {McKernan},
  \& {O'Shaughnessy}}]{2019PhRvL.123r1101Y}
---. 2019{\natexlab{c}}, \prl, 123, 181101,
  \dodoi{10.1103/PhysRevLett.123.181101}

\end{thebibliography}
\end{document}